\def\simless{\mathbin{\lower 3pt\hbox
             {$\rlap{\raise 5pt\hbox{$\char'074$}}\mathchar"7218$}}}    

\def\simmore{\mathbin{\lower 3pt\hbox
             {$\rlap{\raise 5pt\hbox{$\char'076$}}\mathchar"7218$}}}    

\documentclass[structabstract]{aa}

\usepackage{graphicx}
\begin{document}
\title{
Cyclotron line formation by reflection on the surface 
of a magnetic neutron star
}

\subtitle{}

\author{
N. D. Kylafis\inst{1,2} 
J. E. Tr\"{u}mper\inst{3}
\and
N. A. Loudas\inst{1,2}
}

\institute{
University of Crete, Department of Physics \& Institute of
Theoretical \& Computational Physics, 70013 Heraklion, Greece\\
\and 
Institute of Astrophysics,
Foundation for Research and Technology-Hellas, 71110 Heraklion, Crete, Greece\\
\and
Max-Planck-Institut f\"{u}r extraterrestrische Physik, 
Postfach 1312, 85741 Garching, Germany\\
}

\date {Received ; accepted }


\abstract 
{
Accretion onto magnetic neutron stars results in X-ray spectra that often
exhibit a cyclotron resonance scattering feature (CRSF) 
and, sometimes, higher harmonics of it.  
Two places are suspect for the formation of a CRSF:  the surface of the
neutron star and the radiative shock in the accretion column.
}
{
Here we explore the first possibility: reflection at the neutron-star
surface of the continuum produced at the radiative shock.  It has been
proposed that for high-luminosity sources,
as the luminosity increases, the height of the radiative
shock increases, thus a larger polar area is illuminated,
and as a consequence the energy of the CRSF decreases because the dipole
magnetic field decreases by a factor of two from the pole to the equator.
This model has been specifically proposed to explain the observed 
anticorrelation of the cyclotron line energy and luminosity of the 
high-luminosity source V 0332+53.
}
{
We used a Monte Carlo code to compute the reflected spectrum from the 
atmosphere
of a magnetic neutron star, when the incident spectrum is a power-law one. We 
restricted ourselves to cyclotron energies $\ll m_ec^2$ and used 
polarization-dependent scattering cross sections, allowing for 
polarization mode change.
}
{
As expected, a prominent CRSF is produced in the 
reflected spectra if the incident photons are in a pencil beam, which
hits the neutron-star surface at a point with a well-defined magnetic field
strength.  However, the incident beam from the radiative shock has a finite 
width and thus various magnetic field strengths are sampled.  As a result of 
overlap, the reflected spectra have a CRSF, which is close to that produced
at the magnetic pole, independent of the height of the radiative shock.
}
{
Reflection at the surface of a magnetic neutron star cannot explain
the observed decrease in the CRSF energy 
with luminosity in the high-luminosity X-ray pulsar V 0332+53.
In addition, it produces absorption lines much shallower than the 
observed ones.
}


\authorrunning{Kylafis et al. 2021}
\titlerunning{Cyclotron line formation} 

\maketitle


\section{Introduction}

Cyclotron lines observed in the X-ray spectra of accreting neutron stars 
provide direct information for the magnetic field strength in these compact 
objects. The first cyclotron line was discovered in Hercules X-1 
(Tr\"{u}mper et al. 1977, 1978). 
Today we know about 36 objects showing electron 
cyclotron lines, sometimes with harmonics (Staubert et al. 2019).
They are sometimes called cyclotron resonance scattering features (CRSFs).

Despite the fact that a large body of observational data has been collected 
from these sources, it is still not clear where the CRSFs are produced. 
It has been generally assumed that the cyclotron lines are produced at 
the radiative shock (Basko \& Sunyaev 1976) in the accretion column above 
the surface of a magnetic neutron star, but no calculation has been performed 
so far for the formation of a cyclotron line in a radiative shock. 
Typical calculations involve CRSF formation in a slab illuminated from one 
side (Ventura et al. 1979; Nagel 1981; Nishimura 2008; Araya \& Harding 1999, 
2000; Schoenherr et al. 2007). A slab, however, is significantly different 
from a radiative shock in an accretion column, on one side of which there 
is supersonic, free-falling matter and on the other subsonic, thermal plasma. 
In such a situation, both the power-law continuum and the cyclotron line 
are produced by a first order Fermi process at the shock. In other words, 
bremsstrahlung photons from below the shock criss-cross the shock several 
times and, as a result, the in-falling matter is slowed down and a 
power-law hard X-ray spectrum is produced. This has been demonstrated by 
Kylafis et al. (2014), but no resonant scattering cross section was taken 
into account there. A calculation with resonant scattering is currently 
underway.

For five of the 36 cyclotron line sources, significant correlations of 
the line energy with X-ray luminosity have been found by long-term 
observations with different X-ray satellites.  The line energy $E_c$ is 
positively correlated with X-ray luminosity $L_x$, when the luminosity 
is relatively low ($L_x \simless 10^{37}$ erg s$^{-1}$). 
The following four sources of this type are known: 
Hercules X-1 (Staubert et al. 2007; Klochkov et al. 2011), 
GX 304-1 (Klochkov et al. 2012; Malacaria et al. 2015; Rothschild et al. 2017),
A 0535+26 (Klochkov et al. 2011; Sartore et al. 2015), and 
Vela X-1 (F\"{u}rst et al. 2014; La Parola et al. 2016). 
The only object which has been found to show a secure negative correlation 
is V 0332+53 (Mowlavi et al. 2006; Tsygankov et al. 2010; 
Vybornov et al. 2017), 
which has a rather large luminosity (up to $1.5 \times 10^{38}$ erg s$^{-1}$). 
This source shows a transition to a positive correlation at a luminosity of 
$\sim 10^{37}$ erg s$^{-1}$
(Doroshenko et al. 2017; Vybornov et al. 2018).

For an explanation of these correlations the work of Basko \& Sunyaev (1976) 
has been instrumental. It predicts that below a critical luminosity 
$L_x \sim 10^{37}$ erg s$^{-1}$, the infalling matter in the accretion column 
is stopped close to the neutron-star surface by a radiative shock; whereas for 
larger luminosities ($L_x \simmore 10^{37}$ erg s$^{-1}$), the radiative 
shock rises to a height $H$, which is proportional to luminosity. 

Several explanations have been put forward for the positive correlation 
(c.f. Mushtukov et al. 2015 or Mukherjee \& Bhattacharya 2012).  
The negative correlation can be explained qualitatively by 
the decrease in the magnetic field, if, in the high luminosity case, the 
radiative shock rises with luminosity. However, it has been pointed 
out by Poutanen et al. (2013) that in a dipole field, the predicted rate 
of change of $B$ (and $E_c$ ) with $H$ (and $L_x$) is much larger than 
the observed one. As a remedy, the authors propose that the observed 
cyclotron line is not produced in the accretion shock, but from reflection 
of its X-ray beam in the atmosphere of the polar cap. 
Since with increasing $H$ the illuminated polar cap area is increased 
and the strength of the dipole field decreases (up to a factor 2) 
with distance from the magnetic pole, a modest decrease in the cyclotron 
line energy with luminosity is expected. 
In the following, we check on this by means of Monte Carlo simulations. 
In \S~2 we present our model,
in \S~3 we describe briefly our Monte Carlo code,
in \S~4 we show the results of our calculations,
in \S~5 we compare our results with the observations of V 0332+53 by
Doroshenko et al. (2017), which cover a wide range of luminosities,
and in \S~6 we comment on our results and draw our conclusions.

\section{The model}

With the aid of a Monte Carlo code, we study the reflection of continuum
X-ray photons at the atmosphere of a magnetic neutron star.  The photons 
are emitted at the radiative shock, they have
a power-law distribution of energies, and one can think of them as
being emitted at height $H$ above the magnetic pole in the form of a 
wide beam.  
The larger $H$ is, the larger the polar area
on the neutron star that gets illuminated, and the larger the variation
of the magnetic field strength in this area.

It is generally assumed that the beam that emerges from the radiative shock
is directed mainly toward the stellar surface, due to relativistic beaming
(Kaminker et al. 1976; Mitrofanov \& Tsygan 1978; Poutanen et al. 2013).
This, however, is not correct, because it assumes that the escaping photons
from the shock have had their last scattering with a freely falling electron, 
while in a radiative shock, 
it is equally likely that the last scattering is due either
to an in-falling electron above the shock or to a thermal electron below
the shock.  It is well accepted observationally that a fan beam is 
emitted at the shock, but its shape is unknown.  For this reason, 
in our calculations below, and for purposes of demonstration, 
we assume beaming functions (Tr\"{u}mper et al. 2013) of the form
$$
dN/d\theta^\prime \propto \sin^m \theta^\prime,
\eqno(1)
$$  
where $\theta^\prime$ is the polar angle of the direction of the photon
(see Fig. 1),
$m=1$ for an isotropic distribution, and $m=2$ or $3$ for strong
beaming.  
The polar angle $\theta^\prime$ and the angle $\alpha$ between
the photon momentum and the infalling electron velocity are related by 
$\theta^\prime + \alpha = \pi$.  
We have also examined the downward beam distribution of Lyubarskii \& 
Sunyaev (1988) and have found that our conclusions are unaffected.
It is easy to verify that, for angles $\theta^\prime$ such that the photons
encounter the neutron-star surface, the downward 
beam distribution of Lyubarskii \& Sunyaev 
(1988) falls between the $m=1$ and $m=3$ beaming functions.

\begin{figure}[h]
\centering
\includegraphics[angle=0,width=10cm]{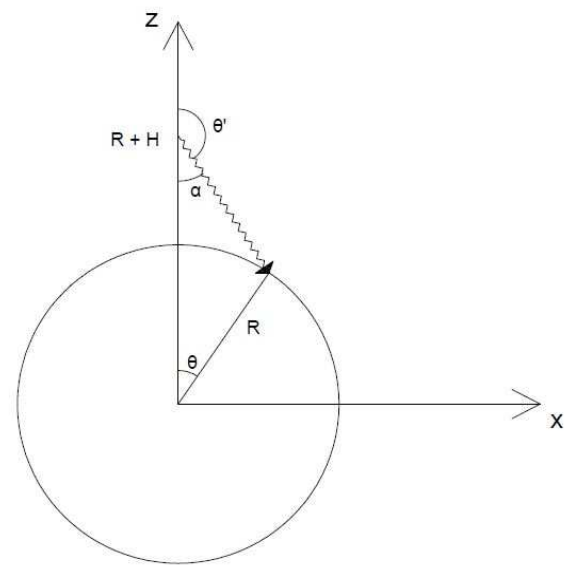}
\caption{
Schematic of a photon emitted at the radiative shock, at height $H$ above
the surface of a neutron star of radius $R$, and hitting the surface.
The polar angle of the photon is $\theta^\prime$ and the corresponding
one at the point of incidence is $\theta$.  The relation between $\theta$
and $\theta^\prime$ is derived in the Appendix.
}
\label{Fig1}
\end{figure}

We assume a dipole magnetic field $\vec B$ with its dipole moment 
$\vec m = m \hat k$ along the $z$ axis.  The vector $\vec B$ on the surface
of the neutron star of radius $R$ is given by
$$
\vec B = B_0 
\left[ 3(\hat m \cdot \hat r)\hat r -\hat m \right],
\eqno(2)
$$
where 
$B_0$ is the magnetic field at the equator of the neutron star
and twice this at its pole,
$\hat r$ is the radial unit vector,
and $\hat m = {\vec m}/m = \hat k$ 
is the unit vector along $\vec m$.
For a point in the $xz$ plane, on the surface of the neutron star,
we write $\hat r = \cos\theta \hat k + \sin\theta \hat i$, where $\theta$
is the polar angle, that is the angle between the radial vector and the $z$
axis, and eq. (2) becomes
$$
\vec B(\theta) = B_0 \left[
(3 \cos^2\theta -1) \hat k + 3 \cos\theta \sin \theta ~ \hat i
\right],
\eqno(3)
$$
The strength of the magnetic field on the surface of the neutron star
at polar angle $\theta$ is
$$
B=B_0 (1+3\cos^2\theta)^{1/2}.
\eqno(4)
$$
The possibility of an off-center dipole exists, but it is beyond
the present paper to investigate it.  We have restricted ourselves to
a centered dipole, as Poutanen et al. (2013) did.

Consider a photon emitted at the shock, at height $H$ 
above the neutron-star surface, with direction
$\hat n = \cos\theta^\prime \hat k + \sin\theta^\prime \hat i$, in the $xz$
plane.  For angles $\theta^\prime > \pi/2$, 
such that $\sin\theta^\prime < R/(R+H)$,
the photon will hit the neutron-star surface at a point $(x, z)$, which is 
the intersection of the straight line path of the photon and the circle
$x^2 + z^2 = R^2$.  The polar angle $\theta$ of this point $(x, z)$,
which will give the
local magnetic field strength $B(\theta)$ (see eq. 4), is determined from
either $\cos\theta = z/R$, $z>0$ or $\sin\theta = x/R$, $x>0$
(see Appendix).  
We do not take into account gravitational bending, 
for two reasons.  First, it is negligible.  Using eqs. (B8) and (B6) of
Poutanen et al. (2013) and our Appendix, we have found that for $H=10$ km
and $m=1$ the magnetic field strength at the circumference of the 
illuminated polar cap is $B=0.69 B_0$ for the Newtonian case and $B=0.62 B_0$
when gravitational bending is taken into account.  A $\sim 10\%$ lower 
magnetic field has no effect in our conclusions.  Second, gravitational
bending brings the emitted photons at the radiative shock
closer to the magnetic pole, and thus strengthens
our main conclusion.

If $\hat n = u \hat i + v \hat j + w \hat k$ is the direction of a photon,
then the angle of incidence with respect to the local magnetic field
$\vec B(\theta)$, at the first or any subsequent scattering, is determined from
$$
\cos\chi = \hat n \cdot { {\vec B(\theta)} \over {B(\theta)} }.
\eqno(5)
$$
Here it is tacitly assumed that the region on the neutron-star surface sampled
by the photon before it escapes 
is very small, i.e., the mean free path is much less than $R$, 
and $\vec B$ is constant in this region.

For a proper treatment of the radiative transfer, one needs to consider 
the complete, Quantum Electrodynamic, Compton cross sections
(Herold 1979; 
Daugherty \& Harding 1986; Harding \& Daugherty 1991;
Sina 1996; Gonthier et al. (2014); Mushtukov et al. 2016; Schwarm 2017).
However, these 
cross sections are quite cumbersome, because they contain infinite sums, 
and are not very practical.
A significant simplification of the above cross sections 
(though still intimidating) was derived by 
Nobili, Turolla, \& Zane (2008), 
in order to treat resonant scattering of photons
with energy approaching $m_ec^2$, in a magnetic field comparable to
the critical one $B_{cr}=m_e^2 c^3/e\hbar $.  These cross sections 
were computed exactly at resonance, i.e., the Lorentz profile was 
approximated by a delta function.

Since most of the cyclotron lines that have been observed so far
(Staubert et al. 2019) are at cyclotron energies $<< m_ec^2$,
we restrict ourselves 
to resonant scattering of photons with energy $E << m_e c^2$ in a magnetic 
field of strength $B << B_{cr}$. 
For such a case, we showed analytically and numerically 
(Loudas, Kylafis, \& Truemper 2021) that the modified classical cross 
sections given below in eq. (6)
provide accurate representations of the fully
relativistic ones.

For a cyclotron energy 
$E_c(\theta)=\hbar \omega_c(\theta) = \hbar eB(\theta)/m_ec \ll m_ec^2$,
the modified classical polarization-dependent resonant cross sections
are given by (Loudas et al. 2021)
$$
{ {d \sigma_{11}} \over {d\Omega^\prime} } \approx
{ {3 \pi r_0 c} \over 8} 
\left(
\cos^2\chi \cos^2 \chi^\prime ~ L_- + 
{1 \over 2} {B \over {B_{cr}} } ~ L_+
\right),
\eqno(6a)
$$
$$
{ {d \sigma_{12}} \over {d\Omega^\prime} } \approx
{ {3 \pi r_0 c} \over 8} 
\left(
\cos^2\chi ~ L_- + 
{1 \over 2} {B \over {B_{cr}} } \cos^2\chi^\prime ~ L_+
\right),
\eqno(6b)
$$
$$
{ {d \sigma_{21}} \over {d\Omega^\prime} } \approx
{ {3 \pi r_0 c} \over 8} 
\left(
\cos^2 \chi^\prime ~ L_- +
{1 \over 2} {B \over {B_{cr}} } \cos^2\chi ~ L_+
\right),
\eqno(6c)
$$
$$
{ {d \sigma_{22}} \over {d\Omega^\prime} } \approx
{ {3 \pi r_0 c} \over 8} 
\left(
L_- + {1 \over 2} {B \over {B_{cr}} } \cos^2\chi \cos^2\chi^\prime ~ L_+
\right),
\eqno(6d)
$$
where the index 1 (2) stands for the ordinary (extraordinary) mode,
$\chi$ and $\chi^\prime$ are the incident and scattered angles, respectively,
with respect to the local magnetic field,
$r_0$ is the classical electron radius, and $c$ is the speed of light.  
The quantities $L_-$ and $L_+$ are the normalized Lorentz profiles
given by
$$
L_{\pm}= {{\Gamma_{\pm}/2\pi} \over 
{(\omega - \omega_r)^2 +(\Gamma_{\pm}/2)^2} },
\eqno(7a)
$$
where $\omega$ and $\omega_r$ are the photon frequency and the resonant
frequency, respectively, in units of $m_ec^2/\hbar$. The decay widths
$\Gamma_{\pm}$ are given by (Herold, Ruder, \& Wunner 1982)
$$
\Gamma_-={4 \over 3} { {m_ec^2} \over \hbar} \alpha 
{ {B^2} \over {B_{cr}^2} },
\eqno(7b)
$$
$$
\Gamma_+={2 \over 3} { {m_ec^2} \over \hbar} \alpha 
{ {B^3} \over {B_{cr}^3} },
\eqno(7c)
$$
where $\alpha$ is the fine structure constant.
We note that $\Gamma_-$ is equal to the classical 
$\Gamma = (4 e^2 \omega_c^2)/ (3 m_e c^3)$, which accounts for the finite
transition life-time of the excited state (e.g., Daugherty \& Ventura 1978;
Ventura 1979).

The first terms in eqs. (6a) - (6d) are the classical (Thomson) differential
cross sections. 
The second terms, which are proportional to $B/B_{cr}$, are purely
quantum mechanical and are
due to spin flip. It is amazing that just these terms are enough
to convert the classical cross sections into accurate approximations of the
fully relativistic quantum-mechanical ones for $B \ll B_{cr}$.

For the resonant frequency $\omega_r$ 
and the new photon energy in the rest frame of the electron after scattering,
we use eqs. (8) and (6), respectively, of Nobili et al. (2008).

The polarization averaged differential cross section is given by
$$
{ {d\sigma} \over {d\Omega'} } \approx 
{ {3\pi r_0c} \over {16} }(1+\cos^2\chi)(1+\cos^2\chi^\prime)
\left( L_- + {1 \over 2} {B \over {B_{cr}} } L_+ \right).
\eqno(8)
$$
For comparison, we have performed a calculation with this polarization 
averaged differential cross section.

The black-body temperature at the surface of the neutron star is 
typically $\simless 1$ keV, which is much smaller that the cyclotron
energy $E_c = 25$ keV at the pole, that we have considered,  
and one could consider the electrons
at the surface of the neutron star as cold.  However, the temperature
of the electrons in the atmosphere of the neutron star can be up to
an order of magnitude larger than the blackbody temperature 
(Zampieri et al. 1995).  Thus, we have taken into account the 
temperature of the electrons.
A one-dimensional (along the local $\vec B$) nonrelativistic Maxwellian 
is fine for our purposes.

Since the heavy elements on the surface of the neutron star will not
be completely ionized, 
for the absorption cross section, we use the hydrogenic approximation
(Bethe \& Salpeter 1957) 
$$
\sigma_a= 
{ {6.3 \times 10^{-18}} \over Z^2 }
(I_Z/E)^3 ~ {\rm cm}^2 ,
\eqno(9)
$$
where $I_Z=13.6 Z^2$ eV is the ionization potential
and we assume that the predominant element is Carbon ($Z=6$).
We remark that our conclusions are not affected at all if there are
no heavy elements on the surface of the neutron star.

\section{The Monte Carlo code}

Our Monte Carlo code is similar to previously used codes in our work 
and it is based on Cashwell \& Everett (1959) and Pozdnyakov et al. (1983).
We emit photons toward the neutron star from height $H$,
with a polar angle $\theta^\prime$, either in the form of a delta-function 
distribution or with continuous distributions given by eq. (1) 
or that of Lyubarskii \& Sunyaev (1988).  The directions
of emission are restricted between 
$\theta^\prime = \arcsin[R/(R+H)]$ (tangent) and $\theta^\prime = \pi$ (pole).

We consider a uniform atmosphere at the surface of the neutron star,
with the number density of the electrons ten times larger than the 
corresponding one for absorbers.  This ratio of densities was selected
so that absorption and scattering have comparable mean free paths at 15 keV.  
Below this energy, absorption dominates, while above it scattering dominates.

In our calculations, we consider a value for $B_0$, such that
$E_c=12.5$ keV $= 11.6 (B_0/1 \times 10^{12})$ keV at the equator 
and 25 keV at the pole.  The variation of
the strength of the magnetic field with polar angle $\theta$ is given
by eq. (4). 
Our input spectra have the form of a power law
$$
{ {dN} \over {dE} } = \left( { {E} \over {E_0} } \right)^{- \alpha},
\eqno(10)
$$
where $E_0=1$ keV is a reference energy and $\alpha$ is the power-law
index.
For the Monte Carlo runs, we have used a minimum of $10^7$ photons.
For high resolution (e.g., Fig. 7), we have used $10^9$ photons.

\section{Results}

As a test of our Monte Carlo code, and in order to see clearly the 
reflected spectrum, we first use a flat power-law spectrum
as input, i.e., $\alpha=0$.  For the photons to see a specific magnetic
field strength, we send all of them to the pole, i.e., $\theta^\prime = \pi$.
Equally well, we could have sent them to the tangent 
($\theta^\prime = \arcsin[R/(R+H)]$) or any intermediate direction.
The temperature of the electrons is taken equal to zero, as a benchmark.
In Fig. 2, we show our results.  The black dashed line
represents the input spectrum and the black crosses give the reflected
spectrum.  However, the observed spectra consist of two components:  
the photons that are emitted at the radiative shock and come directly
to the observer and those that come to the observer after being
reflected at the neutron-star surface.  The relative strength of these
two intensities is difficult to compute, because it
depends on two angles, both unknown:  the inclination of the spin 
axis of the neutron star and the angle between the spin and the magnetic 
axes.  For most combinations of these two angles, the direct spectrum is 
more than 50\% of the spectrum of the photons emitted at the radiative 
shock. Also, the larger the height $H$ of the radiative shock, the larger 
the direct component. Therefore, in Fig. 2, we show mixtures of 50\% 
direct and 50\% reflected (blue stars) and 70\% direct and 30\% reflected 
(red pluses) as representative examples. 

\begin{figure}[h]
\centering
\includegraphics[angle=-90,width=10cm]{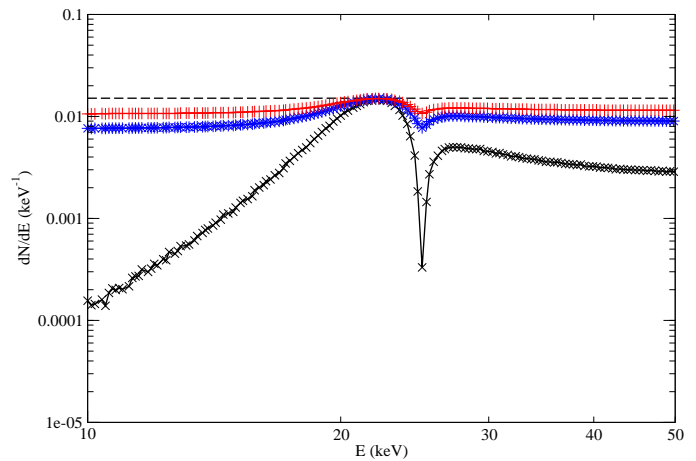}
\caption{
Reflected spectrum $dN/dE$ in all directions (black crosses) in
arbitrary units, as a function of energy $E$ in keV.  The
input spectrum is shown as a black dashed line.  The other two lines
show mixed spectra:  50\% direct and 50\% reflected (line with blue stars) and
70\% direct and 30\% reflected (line with red pluses).
}
\label{Fig2}
\end{figure}

The reflected spectrum in Fig. 2 (black crosses) has the 
expected characteristics.
First, the input photons near resonance have a small mean free path
in the neutron-star atmosphere and, as a result,
a good fraction of them get reflected.  However, with every
scattering these photons lose energy due to electron recoil.  Thus,
the reflected photons appear at energies below the resonance.
Second, the resonant input photons that do not get reflected, but
instead go deeper into the neutron-star atmosphere, face a large
optical depth for escape and thus they escape only in the line
wings.  Similarly, photons with energy somewhat larger than $E_c$ 
that go deep in the atmosphere, loose energy with every scattering and can 
escape, if they are not absorbed, only in the line wings.
Thus, an absorption feature is generated at the 
resonance.  However, since the atmosphere of the neutron star
near resonance is an essentially pure scattering one, the number of photons
is conserved.  The photons that are
missing in an absorption feature, must appear
as an emission feature at lower energies, due to electron recoil.

At low energies, the reflected spectrum is significantly reduced 
due to absorption (see eq. 9). At high energies, the reduction 
of the reflected spectrum is due to down-scattering, followed by absorption.

Furthermore, we see in Fig. 2 that,
despite the fact that the reflected spectrum
has a sharp and prominent absorption feature, this feature is not so
prominent when the reflected spectra are mixed with nonreflected
ones, i.e., with input spectra that reach the observer directly.
Naturally, the larger the direct component in the mixture, the
shallower the absorption features and vice versa.

We repeated the calculation of Fig. 2, but with the polarization averaged
differential cross section of eq. (8), instead of the mode-dependent ones of 
eqs. (6a - 6d).  The results are indistinguishable from those of Fig. 2.

In Fig. 3, we show the results of the same calculation, but with 
the thermal motion of the electrons taken into account.  
We consider values of $kT_e$ equal to 
1 keV (line with black stars), 
3 keV (line with blue circles), 
and 10 keV (line with red diamonds).  
The input spectrum is shown as a black dashed line.

\begin{figure}[h]
\centering
\includegraphics[angle=-90,width=10cm]{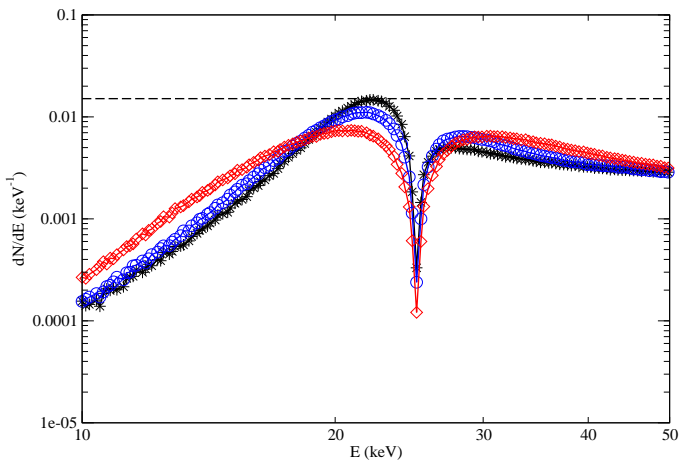}
\caption{
Reflected spectrum $dN/dE$ in all directions in
arbitrary units, as a function of energy $E$ in keV
for $kT_e = 1$ keV (line with black stars), 
$kT_e = 3$ keV (line with blue circles), 
and $kT_e = 10$ keV (line with red diamonds).  The
input spectrum is shown as a black dashed line.  
}
\label{Fig3}
\end{figure}

Contrary to our intuition regarding the thermal
broadening of an absorption line, where the Lorentz profile is blue-shifted
and red-shifted by the thermal motion of the absorbers, here the thermal
motion of the electrons has a small and rather undetectable effect on 
the CRSF.  The explanation is as follows.

Independent of the temperature of the electrons
(assuming $kT_e \simless 10$ keV, as it is thought appropriate
for a neutron star atmosphere), the discussion 
that we have given above
for the formation of the absorption feature still holds.  Photons at
resonance cannot escape from deep in the atmosphere
unless they change their energy and escape in 
the line wings.  If the electrons are cold, resonance photons lose
energy with every scattering and escape in the red wing.  However, if the
electrons are hot, they can give energy to the photons, which then
escape in the blue wing.  The hotter the electrons, the larger the blue
wing.  This is exactly what is seen in Fig. 3. 

In the subsequent calculations and for concreteness, we take $kT_e = 1$ keV,
though the spectra are indistinguishable from those with $kT_e = 0$.  
We stress that our conclusions are independent of the temperature of the
electrons.

In Fig. 4, we show the dependence of the reflected spectrum on the direction
of escape.  The incident photons are again in the form of a pencil beam 
toward the pole.
The line with the black circles gives the spectrum of the
reflected photons that escape with angle $\chi^\prime$ with respect
to the local 
$\vec B=2B_0 \hat k$, such that $0 \le \cos \chi^\prime < 0.33$.  
The line with the blue diamonds corresponds to
$0.33 \le \cos \chi^\prime < 0.66$
and the line with the red triangles corresponds to
$0.66 \le \cos \chi^\prime \le 1.0$.
The three reflected spectra are very similar.

\begin{figure}[h]
\centering
\includegraphics[angle=-90,width=10cm]{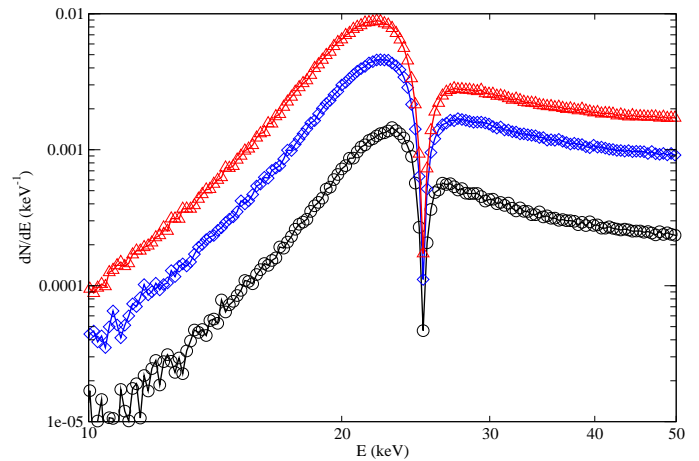}
\caption{
Reflected spectrum $dN/dE$ in three direction bins, as follows: 
$0 \le \cos \chi^\prime < 0.33$ (line with black circles),
$0.33 \le \cos \chi^\prime < 0.66$ (line with blue diamonds), and
$0.66 \le \cos \chi^\prime \le 1.0$ (line with red triangles). 
The angle $\chi^\prime$ is measured with respect to $\vec B$
at the point of incidence.
}
\label{Fig4}
\end{figure}

In Fig. 5, we use as input spectrum a steep power law with index 
$\alpha=5$, as suggested by the spectra in Tsygankov et al. (2006),  
Cusumano et al. (2016), and Doroshenko et al. (2017). 
Also, instead of $dN/dE$ we show $E^2 ~ dN/dE$ versus
energy.  The angle of incidence is
$\theta^\prime=\pi$ and the symbols
are the same as in Fig. 2.  The reflected spectrum has a prominent
absorption feature, but in the mixed spectra (lines with stars and pluses) 
it is not as prominent.

\begin{figure}[h]
\centering
\includegraphics[angle=-90,width=10cm]{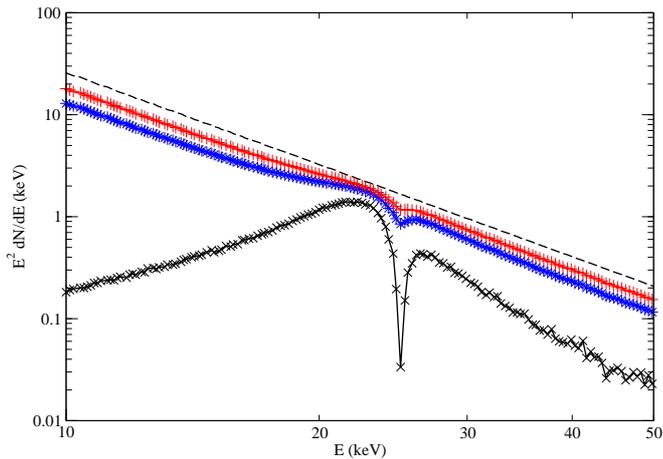}
\caption{
Reflected spectrum in all directions in $E^2 dN/dE$ form (crosses), with 
arbitrary units, as a function of energy $E$ in keV.  
The input spectrum is shown as a dashed line.  The other two lines
show mixed spectra:  50\% direct and 50\% reflected (stars) and
70\% direct and 30\% reflected (pluses).  The angle $\theta^\prime = \pi$.
}
\label{Fig5}
\end{figure}

In all of the above cases, we have considered a 
delta-function beam from the shock to the magnetic pole.  
However, as we have discussed in sections 2 and 3, the beam that is
emitted at the shock is wide and it
samples various magnetic field strengths on the surface
of the neutron star.  For a rather large shock height $H=10$ km and 
$R=12.5$ km, the beam extends from 
$\theta^\prime = \arcsin[R/(R+H)]= 146.3$ degrees (tangent) to 
$\theta^\prime = 180$ degrees (pole). 
At $\theta^\prime = 146.3$ degrees, the photon hits the neutron-star surface 
at a polar angle (see Fig. 1)
$\theta = 53.4$ degrees (see Appendix).  
The local magnetic field strength is
$B= 1.44 B_0$, with a corresponding cyclotron energy
$E_c= 17.97$ keV.

In Fig. 6, we show the reflected spectra from a number of delta-function
beams from $\theta^\prime = 147$ degrees (grazing incidence) 
to $\theta^\prime = 180$ degrees (vertical incidence)
for a shock height $H=10$ km.
All delta-function beams have the same number of photons and their spectrum
is that of the black dashed line.
The red dotted line corresponds to $\theta^\prime = 147$ degrees, with the 
local cyclotron energy $E_c = 19.63$ keV at the point of incidence.
The green dot - dashed line corresponds to $\theta^\prime = 149$ degrees
(local $E_c= 21.34$ keV),
the blue double-dot - dashed line corresponds to $\theta^\prime = 152$ degrees
(local $E_c= 22.59$ keV),
the magenta double-dash - dotted line corresponds to 
$\theta^\prime = 155$ degrees (local $E_c= 23.33$ keV),
the black solid line corresponds to $\theta^\prime = 160$ degrees
(local $E_c= 24.09$ keV), and
the brown long-dashed line corresponds to $\theta^\prime = 172$ degrees
(local $E_c= 24.88$ keV),
As expected, the energy of the absorption feature increases 
with increasing $\theta^\prime$ from 17.97 keV to 25 keV.  

\begin{figure}[h]
\centering
\includegraphics[angle=-90,width=10cm]{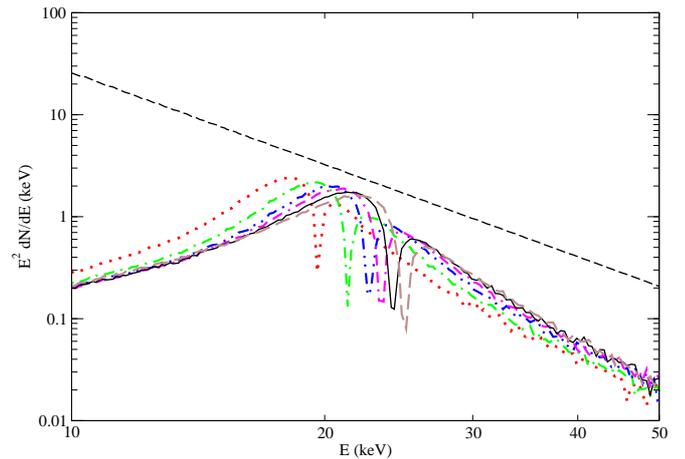}
\caption{
Reflected spectra in all directions in $E^2 dN/dE$ form, with 
arbitrary units, as a function of energy $E$ in keV.  
For all cases, the input spectrum is given by the black dashed line.
The red dotted line corresponds to $\theta^\prime = 147$ degrees,
the green dot - dashed line corresponds to $\theta^\prime = 149$ degrees,
the blue double-dot - dashed line corresponds to $\theta^\prime = 152$ degrees,
the magenta double-dash - dotted line corresponds to $\theta^\prime = 155$ 
degrees,
the black solid line corresponds to $\theta^\prime = 160$ degrees, and
the brown long-dashed line corresponds to $\theta^\prime = 172$ degrees.
The corresponding energies of the cyclotron absorption features 
are given in the text.  The shock height is $H=10$ km.
}
\label{Fig6}
\end{figure}

Two additional things are worth noticing in Fig. 6.
1) The absorption features are weaker when the photons have a 
grazing incidence in the atmosphere ($\theta^\prime=147$ degrees)
than when they have a nearly perpendicular one ($\theta^\prime=172$ degrees).
This is understood, because at grazing incidence the photons see a
smaller absorption depth for escape than at normal incidence.
2) The absorption feature at grazing incidence 
($\theta^\prime=147$ degrees) falls on top of the emission feature 
at incidence at $\theta^\prime=149$ degrees, and similarly for
subsequent incidence angles.  This has the important implication that
when a beam falls on the neutron star, the addition of the reflected
spectra, properly weighted with the angular distribution of the beam,
will wash out the imprints of weaker magnetic fields from the 
low-magnetic-latitude part of the polar cap.
We have found that this is indeed the case
{\it
and this is the main point of our paper.
}

In Fig. 7, we show 
the reflected spectrum (red dot-dashed line) 
from an isotropic beam ($m=1$ in 
eq. 1) and similarly (blue solid line)
from a strongly beamed distribution ($m=3$ in eq. 1). 
In both cases, the input spectrum is given by the black dashed line. The
shock height is $H=5$ km, which implies that, on the illuminated surface
of the neutron star, $E_c$ ranges from 20.0 to 25.0 keV.  
In both cases, the absorption feature from a
broad beam is naturally wider than the corresponding one from a pencil
beam (see Fig. 6).  The minima, though, of the absorption features are 
at 24.7 keV for the $m=1$ case and at 24.5 keV for the $m=3$ case, 
which are very close to the cyclotron energy at the pole.
The contribution of the lower magnetic 
latitudes is nearly washed out.
The strongly beamed distribution ($m=3$) produces an absorption
feature (blue solid line in Fig. 7), which has slightly lower 
energy than that of the 
isotropic distribution (red line in Fig. 7, $m=1$).  This is because, in the
$m=3$ case, most of the photons hit the neutron-star surface at low
magnetic latitudes and very few near the pole.
We note that Fig. 7 shows only the reflected spectrum.  The observed 
spectrum, however, is a combination of direct and reflected spectra, 
which makes the absorption features shallower (see Fig. 5).

\begin{figure}[h]
\centering
\includegraphics[angle=-90,width=10cm]{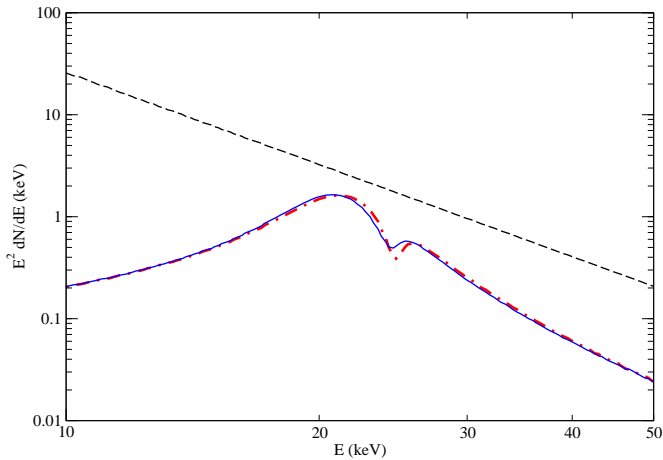}
\caption{
Reflected spectra in all directions in $E^2 dN/dE$ form, with 
arbitrary units, as a function of energy $E$ in keV.  
The red dot-dashed line corresponds to a beam with $m=1$, while the 
blue solid line
to a beam with $m=3$.  The shock height is $H=5$ km
and the input spectrum is given by the black dashed line.
}
\label{Fig7}
\end{figure}

A shock height $H=5$ km is rather large.  We used it only 
for reasons of demonstration.  In Fig. 8, we show the reflected
spectrum of a beam with $m=1$ and a reasonable 
shock height $H=2$ km.  A beam with $m=3$ produces nearly identical 
results.  The absorption 
feature occurs at the cyclotron energy corresponding to the magnetic 
field at the pole and, 
as expected, it is narrower than that of Fig. 7,
because a narrower range of magnetic field strengths is sampled
($22.74 \le E_c \le 25.0$ keV).
In the mixed spectra (stars and pluses), however, the absorption feature 
is less prominent.

\begin{figure}[h]
\centering
\includegraphics[angle=-90,width=10cm]{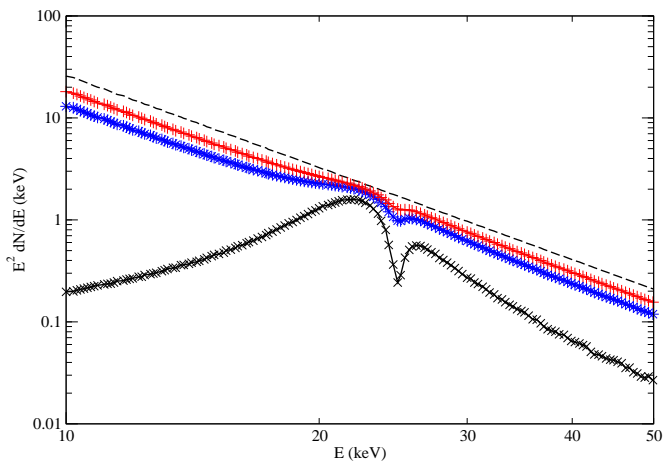}
\caption{
Reflected spectrum (crosses) in all directions in $E^2 dN/dE$ form, with 
arbitrary units, as a function of energy $E$ in keV.  
The photons are emitted in a beam with $m=1$. The shock height is $H=2$ km.
The other two lines show mixed spectra:  50\% direct and 50\% reflected
(stars) and 70\% direct and 30\% reflected (pluses).
The input spectrum is given by the black dashed line.
}
\label{Fig8}
\end{figure}

\section{Comparison with observations}

With regard to the change of the cyclotron-line energy with luminosity, 
the most remarkable X-ray pulsar to date is V 0332+53. It was demonstrated
by Doroshenko et al. (2017) that, in its 2015 and 2016 outbursts, the 
cyclotron-line energy exhibited both a correlation 
(at low luminosities) and an anticorrelation (at high luminosities) with
luminosity.  Both, the correlation and the anticorrelation have been
seen in other sources (for a review see Staubert et al. 2019), but todate
this is the only source that exhibits both.  

During the decay of the luminosity, the cyclotron-line energy increased 
from 27.9 keV to 30.4 keV.  This is approximately a 9\% increase.

To compare with these observations, we have run models with the shock
height at $H=10, 8, 5, 3$, and $1$ km to mimic the decrease in the 
observed luminosity.  We have also used a beam with $m=3$, which is
the most favorable to exhibit a variation of the cyclotron-line energy
with shock height or equivalently luminosity.
In Fig. 9 we show our results.

\begin{figure}[h]
\centering
\includegraphics[angle=-90,width=10cm]{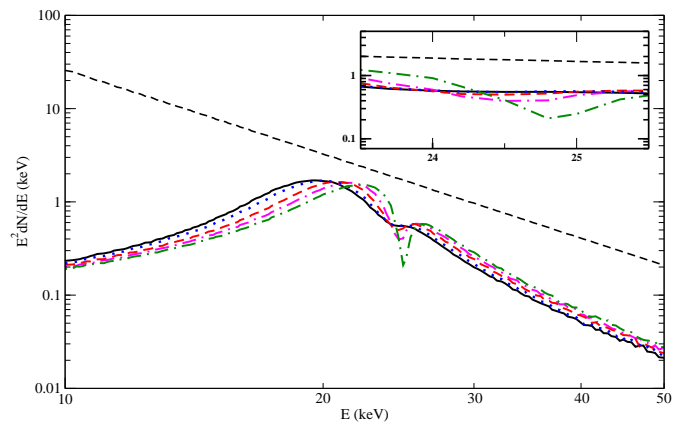}
\caption{
Reflected spectra in all directions in $E^2 dN/dE$ form, with 
arbitrary units, as a function of energy $E$ in keV
for different shock heights:  $H=10$ km (black solid line), $H=8$ km 
(blue dotted line), $H=5$ km (red dashed line), $H=3$ km 
(magenta dot-dashed line), and
$H=1$ km (green dash - dotted line). The dashed straight line is 
the input spectrum. 
The photons are emitted in a beam with $m=3$. In the inset, we zoom
into the absorption feature, with the horizontal axis linear.
}
\label{Fig9}
\end{figure}

In varying the shock height by an order of magnitude, the cyclotron line 
varies from $24.5\pm 0.1$ keV (the uncertainty is due to binning) 
to 25 keV, i.e., only about 0.2\%.  This is practically
unobservable.  

We remark that had we used a downward beam, most
of the photons would hit the neutron-star surface near the magnetic 
pole and the variation of the cyclotron-line energy would be even
smaller.

\section{Discussion and conclusions}

We have studied reflection of continuum 
X-ray photons from the surface of a magnetic neutron star.
With a Monte Carlo code, we have simulated photons emitted from a
radiative shock at height $H$ above the surface of the neutron star,
on the magnetic axis.
The photons are either unidirectional or are in beams with specific
distributions: isotropic, fan beam, or downward beam.

When the photons have a specific direction, they impinge in a polar ring
on the neutron-star surface, that has a specific magnetic field.  The lower
the magnetic latitude of the ring, the lower the strength of the local dipole
magnetic field.  In such a case, the reflected spectrum exhibits a prominent
cyclotron absorption line at the local cyclotron energy.
Naturally, if the photons are emitted in 
a beam, a range of magnetic fields is sampled and the absorption line
is wider.

If we could see only the reflected spectrum, these absorption features
would be easily detectable.  However, the observed X-ray spectra are a 
mixture of direct photons from the radiative shock and reflected
ones.  We have found that, for reasonable mixtures of direct and 
reflected photons, 
which depend on the inclination angle of the spin axis as well as
on the angle between the magnetic and the spin axes (both unknown),
the cyclotron features become much weaker than the ones
observed (see, e.g., Fig. 3 of both Tsygankov et al. 2006
and Doroshenko et al. 2017).

Another more serious finding, 
that speaks against the idea of cyclotron line formation
at the surface of magnetic neutron stars, is that the reflected spectra
are essentially unaffected by the height of the radiative shock.
This is contrary to intuition, where one would expect an anticorrelation
between the luminosity of the source, and therefore the height of the 
shock, and the cyclotron absorption feature (Poutanen et al 2013).
The reason for this is the following.  

Consider photons at grazing incidence to the neutron-star surface.  These
photons impinge on the furthermost ring of the polar cap, which has
the lowest magnetic field in the polar cap.
Reflection of these photons
creates a) an absorption feature at the local cyclotron energy and b)
an emission feature below this.
Consider now photons impinging on the adjacent ring, inside the first one.
The strength of the magnetic field here is larger than that of the first ring.
Thus, the emission feature at the local cyclotron 
energy falls on top of
the absorption feature of the first ring, and 
the absorption feature is nearly washed out.  The same
thing happens with all inner rings.
Since, in the illuminated polar cap, the magnetic field strength 
varies continuously from the circumference to the magnetic pole, 
illumination of the polar cap results in
a broad cyclotron line, like the one shown in Fig. 7.  The magnetic 
field inferred from such a line corresponds to that of the magnetic pole.
We remark that
the distribution of the photons in the beam
is practically irrelevant.
An isotropic beam and a strong fan beam give practically the same reflected 
spectrum.
For this reason, we think that the use of 
gravitational bending will not alter the above picture.

Last, but not least, we remark that a shock height $H=10$ km is 
unreasonably high.  It was used only for the purpose of demonstration.
For more reasonable heights of a few kilometers, 
the variation of the magnetic field in the polar cap is small and the spectra
exhibit practically a broad absorption feature, which, in addition,
is very shallow if the reflected spectra are mixed with direct ones.  
Thus, reflection at the surface of a neutron star is not the
mechanism for cyclotron line formation.

Due to the overlap of the CRSFs at different magnetic-field strengths,
reflection at the surface of a neutron star cannot explain
the anticorrelation of cyclotron line energy and luminosity. 
Several alternative scenarios have been mentioned by 
Becker et al. (2012) Doroshenko et al. (2017), 
and Staubert et al. (2019), but a quantitative explanation of the effect 
is still pending.

\begin{acknowledgements}
N.D.K. would like to thank Sterl Phinney for his notes on the classical
magnetic scattering cross sections.
\end{acknowledgements}

\newpage
\begin{appendix}
\section{Geometry}

We consider a point source on the magnetic axis $z$, at height $H$, above the 
surface of a neutron star.  The source emits a pencil beam of photons 
with direction
$$
\hat n = \cos\theta^\prime \hat k + \sin\theta^\prime \hat i
$$
in the $xz$ plane, with the polar angle $\theta^\prime > \pi/2$ such that 
$\sin\theta^\prime \le R/(R+H)$.  Under these conditions, the pencil beam
hits the neutron-star surface at a point $(x,z)$, which is determined
from the intersection of the straight line
$$
z=\cot\theta^\prime x+ R+H
$$
and the circumference of the circle
$$
x^2+z^2=R^2.
$$
Eliminating $z$ from these two equations we obtain
$$
(1+\cot^2\theta^\prime)x^2+2\cot\theta^\prime(R+H)x+(R+H)^2-R^2=0,
$$
from which we get 
$$
x={ {-b -\sqrt{b^2-4ac}} \over {2a}}>0
$$
and
$$
z=\sqrt{R^2 -x^2} >0,
$$
where
$$
a=1+\cot^2\theta^\prime >0,
$$
$$
b=2\cot\theta^\prime(R+H) <0,
$$
and
$$
c=(R+H)^2-R^2.
$$
The polar angle $\theta$ of the point $(x,z)$ is determined from
$$
\cos\theta=z/R,~~~~0 \le \theta < \pi/2.
$$

\end{appendix}

\end{document}